# Analysis of Longitudinal Changes in Privacy Behavior of Android Applications

ALEXANDER YU, Carnegie Mellon University
YUVRAJ AGARWAL, Carnegie Mellon University
JASON I. HONG, Carnegie Mellon University

Privacy concerns have long been expressed around smart devices, and the concerns around Android apps have been studied by many past works. Over the past 10 years, we have crawled and scraped data for almost 1.9 million apps, and also stored the APKs for 135,536 of them. In this paper, we examine the trends in how Android apps have changed over time with respect to privacy and look at it from two perspectives: (1) how privacy behavior in apps have changed as they are updated over time, (2) how these changes can be accounted for when comparing third-party libraries and the app's own internals. To study this, we examine the adoption of HTTPS, whether apps scan the device for other installed apps, the use of permissions for privacy-sensitive data, and the use of unique identifiers. We find that privacy-related behavior has improved with time as apps continue to receive updates, and that the third-party libraries used by apps are responsible for more issues with privacy. However, we observe that in the current state of Android apps, there has not been enough of an improvement in terms of privacy and many issues still need to be addressed.

CCS Concepts: • **Security and privacy** → **Privacy protections**; • **Human-centered computing** → *Empirical studies in ubiquitous and mobile computing*; *Smartphones*.

## 1 INTRODUCTION

The most prevalent smartphone platform today is Android, with one estimate that it is running on 76% of smartphones worldwide [31]. The official Android app marketplace, known as Google Play, has also grown tremendously and now contains over 2.7 million different apps [3]. The appeal of many of these apps is their customizability and diversity in services provided. However, some of these capabilities, in particular being able to access sensitive sensor data and personal data stored on the device, has raised multiple privacy concerns.

A number of studies have examined privacy concerns surrounding Android apps and potential bad actors in the marketplace from varying perspectives. One area that has been extensively studied in the past is understanding the role that third-party libraries play in privacy issues (e.g. [10, 32]). Another area of work has been investigating abnormal privacy-related behavior in apps. Some systems developed in previous work, such as CHABADA [18], WHYPER [25], and AutoCog [27], did so by comparing permissions requested against the app's description against its actual behavior, while some other works traced the flow of accessed resources [5, 38]. Some other different lines of past work have examined specific parts of privacy (e.g. [28, 29]), the perspective of users on privacy issues [22], and broad surveys of the Android app ecosystem (e.g. [35]).

We further contribute to these works with analyses of the data we have collected on Android apps over the last 10 years. In that time, we have accumulated a significant amount of metadata and number of APKs for apps, with versions spanning across that period of time. We are not the first to analyze apps at scale longitudinally. PlayDrone [34] was one the first systems to index and analyze Google Play at scale, and the analysis conducted by Viennot et al. was over 610,000 apps and spanned the months of April and November in 2013. On the other hand, Ren et al. only analyzed 512 apps but did so over the course of 8 years [29]. Our analyses are conducted at both a large scale and scope of time. The number of apps analyzed ranges from 60,000 to 350,000, depending on the aspect of analysis, and these apps have versions that range from early 2013 to early 2019.

In this paper, we look at general changes and trends in Android apps published on Google Play, with respect to privacy. We set out to use this data to answer two key questions of interest. The first question we consider is how the age of an app, or the amount of time it is in Google Play, affects the problems it has with respect to privacy? This is an important point to consider because apps would ideally improve their behavior with time, especially given previous work identifying various issues. Second, privacy problems within Android apps have also often been attributed to the third-party libraries used. To see if this is still valid, we consider the question of how the changes in an app's privacy-related behavior can be accounted for, between its internal code and the third-party libraries? To investigate privacy-related behavior, we examine the adoption of HTTPS, whether apps scan the device for other installed apps, the use of permissions to request privacy-sensitive data, and the use of various types of unique identifiers. We chose these aspects because of their ubiquity across apps and their ease of analysis at scale.

We observed a few key findings about how privacy has changed for apps over time. First, we found that behavior with respect to privacy has improved with time as apps continue to receive updates from developers. This corroborates the conclusions of previous work that demonstrated that developers considered privacy to be of a lower priority than other factors, such as app functionality and usability [1, 7, 21, 23]. Second, we observed that the third-party libraries used by apps are responsible for more issues with privacy than the internals of the app itself, offering further support for the findings of previous work [6, 8, 10, 32]. In addition, we offered new evidence for this claim with the examination of HTTPS adoption, scanning for other installed apps, and unique identifier usage.

In this paper, our main contributions are: (1) an extensive static analysis of certain key components of privacy-related behavior demonstrated by Android apps over the course of the past few years, (2) our findings with regards to the questions we laid out about the changes in privacy in the analyzed apps, (3) the corroboration of results found by previous works to show that these conclusions are still true in the present.

## 2 RELATED WORK

Third-party libraries and their usage by Android apps have been studied in detail in the past. For example, Ma et al. proposed a tool called LibRadar that quickly and accurately determined the libraries used by an app, which was done by examining the frequency of different API calls in a way that was resilient to obfuscation [24]. Backes et al. also proposed a detection technique that was obfuscation resilient but instead used profiles from a database of libraries and the namespace tree of a Java package. There has also been a large focus on the privacy risk presented by these libraries. In addition to the proposed detection technique, Backes et al. also demonstrated that libraries increase the attack surface on apps by misusing cryptographic APIs. It was also shown that ad libraries in particular posed risks to privacy by checking for privacy-invasive permissions, such as READ_PHONE_STATE [10]. Furthermore, libraries were found to have used permissions beyond what was listed in their respective documentations and often included dangerous permissions such as CAMERA, WRITE_CALENDAR and WRITE_CONTACTS [32].

Another significant area of past work has been investigating abnormal privacy-related behavior in apps and what abnormal behavior entails. The CHABADA [18], WHYPER [25], and AutoCog [27] systems all employed some type of natural language processing technique to analyze the description of an app. They then compared those results against the APIs or permissions used by the app to determine which apps demonstrated abnormal behavior. This conclusion was based on whether or not the app's description provided sufficient reason for why each API or permission was necessary. Other works in this area traced the flow of privacy-sensitive behaviors [5] and the context they are used in [38], then determined whether each flow was abnormal or not.

Some other different lines of past work have examined specific parts of privacy. Ren et al. conducted a longitudinal analysis of different types of personally identifiable information, or PII, leaks, which included analyses of unique identifiers used for tracking and the adoption of HTTPS [29]. Similarly, Razaghpanah et al. also studied the adoption of HTTPS in apps, but studied the implementation and found vulnerabilities in the configurations and APIs used [28].

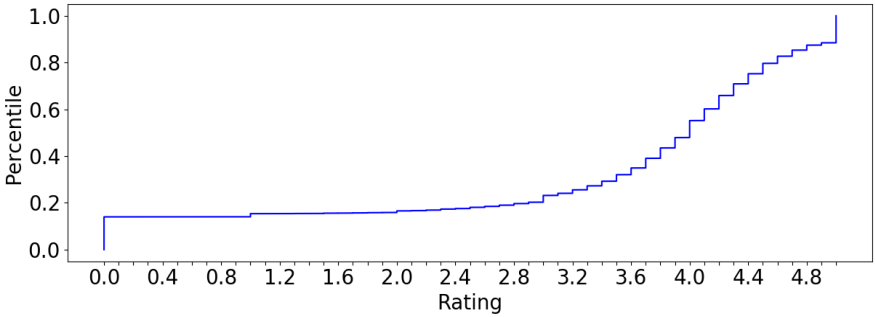

Fig. 1. The distribution of ratings of the apps we have collected shows that the most common ratings are 0 and 5 stars, with over half of the apps having a rating of 4 stars or more.

The novelty of our work presented in this paper lies in that it combines aspects from each of these areas of previous work. We analyze the role of third-party libraries in privacy issues but also compare it against the role played by the internal code of apps. Although we do not look at whether an app's behavior is abnormal or not, we do incorporate the analysis of permissions as one of the aspects of privacy considered. In addition, we include other specific parts of privacy that have been analyzed before, namely the adoption of HTTPS. These are all parts of our work that we present as part of a broader investigation of the trends in privacy.

## 3 DESCRIPTIVE STATISTICS OF OUR DATA SET

As mentioned previously, the data set used for the analyses conducted in this paper was expansive in terms of both the number of apps crawled and the period of time over which the data was collected. Over the course of the past 10 years, we have collected data for almost 1.9 million apps from Google Play. This data set was crawled and scraped from the official store via a Python API, and contains data regarding the app information, developer information, and details about its performance among Android users. Out of these apps, we have downloaded and stored the APKs for 135,536 of them. These apps form the major bulk of our analyses presented in this paper as they required decompiling and parsing the contents of the APK.

We first present some high-level, descriptive statistics about our data set. The 1.9 million apps collected span all periods of time from the last 10 years, from April 27th, 2009 to February 7th, 2019. Of these apps, 251,637 of them were paid and the rest were all free. All of the apps that we stored APKs for were also all free apps. Table 1 shows a breakdown of the apps we have collected by category. As the data shows, games were by far the most common type of app found, which is why we further broke down the categories of those apps into the specific game genre as well. Figure 1 shows the distribution of the user ratings for these apps. The two most common ratings are 0 stars and 5 stars, with 14.0% and 11.6% of apps having those ratings, respectively. Furthermore, 52.1% of the apps we scraped over this period of time had a user rating of 4 stars or more. Finally, Figure 2 shows the distribution of the number of downloads for each of these apps. This showed that the majority of apps do not even surpass 5,000 downloads, with 64.6% of apps falling into this category.

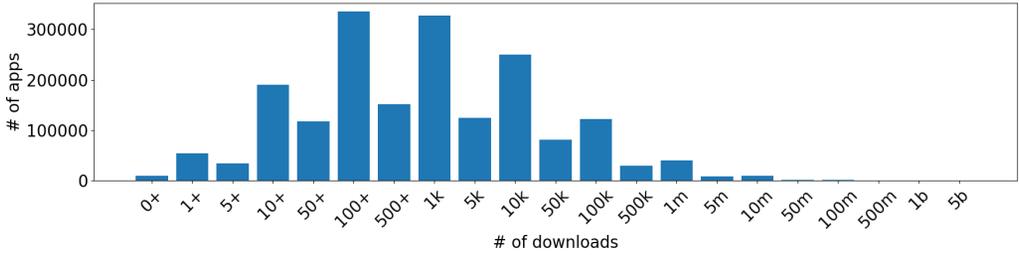

Fig. 2. The distribution of the number of downloads of the apps we have collected demonstrates that a large majority of apps only received less than 5,000 downloads. Only a few percent of apps we crawled had received 1 million or more downloads.

| Category | # of apps |
|---|---|
| Games | 251,688 |
| Entertainment | 142,274 |
| Education | 107,085 |
| Personalization | 105,965 |
| Lifestyle | 102,234 |
| Tools | 92,278 |
| Books & Reference | 82,281 |
| Business | 75,609 |
| Travel & Local | 62,189 |
| Music & Audio | 61,130 |
| News & Magazines | 42,818 |
| Health & Fitness | 40,699 |
| Sports | 38,945 |
| Productivity | 36,418 |
| Communication | 33,437 |
| Social | 30,867 |
| Finance | 28,723 |
| Photography | 24,553 |
| Media & Video | 23,541 |
| Shopping | 22,970 |
| Game - Puzzle | 52,406 |
| Game - Casual | 49,244 |
| Game - Arcade | 40,623 |
| Game - Action | 18,226 |
| Game - Educational | 11,828 |
| Game - Simulation | 10,620 |
| Game - Adventure | 10,606 |

Table 1. This table shows the number of apps in each category from the apps we have crawled. Games apps are by far the most common, with the other popular categories with more than 100,000 apps being entertainment, education, personalization and lifestyle.

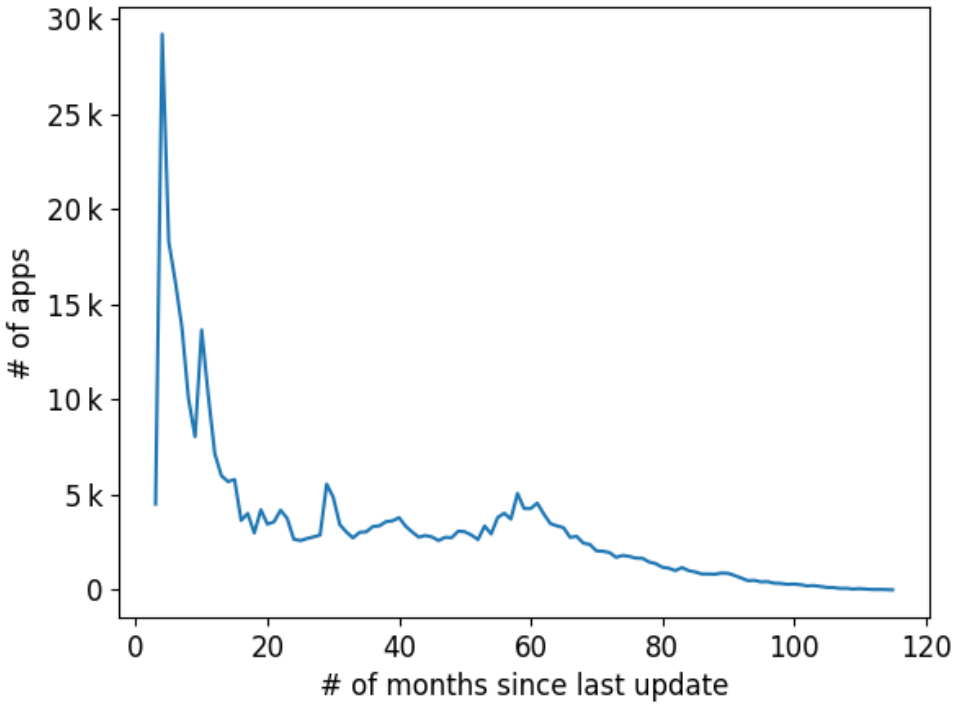

Fig. 3. This graph shows the number of months that have passed since the last update of an app as of 05/08/2019. The data is only shown for apps that were not yet removed at the time of our last crawl in early 2019. We found that a significant portion of apps had received an update within the past year.

Furthermore, a slim percentage of apps, specifically 3.2%, of apps we crawled had received 1 million or more downloads.

Figure 3 shows the distribution of the number of months that have passed since the last update of each app in our data set, assuming that it was not removed from Google Play. We observe that 35.9% of the 364,676 apps represented in this figure had received an update within the past year, as of August 15th, 2019. Although this does not represent a majority of the apps, we believe that this statistic would be higher if our crawls had happened more frequently. This was one of the limitations of our crawl; we will discuss this further in the next section.

## 4 APP ANALYSIS FRAMEWORK

The main goal we had in mind when building out this app analysis framework was to allow the research community to access the data collected for various types of research regarding Android apps in the future. The framework was also designed to be able to operate at scale, and be extensible as more and more functionality is desired. The framework consists of two main components, a pipeline that crawls and analyzes data, and an API layer that exposes the data. The framework resides on a virtual machine on a compute server shared among our lab. The VM we use is currently allotted 36 cores and 132 GB of memory. Any large objects, such as APKs and pre-decompiled APKs, are kept on a storage server that is mounted as a NAS and has 36 TB of storage capacity.

## 4.1 Data Pipeline

The data pipeline is responsible for the entire process of getting data, from crawling for new apps and to analyzing the downloaded APK. It consists of five main modules, each of which is also extensible via our mechanism for dynamically loading plugins that follow a given pre-defined structure. Our code for the pipeline can be found on Github at https://github.com/CMUChimpsLab/playstore-scraper.

*4.1.1 Crawler.* The crawler portion of the pipeline crawls for apps in two different ways. The first way is just to crawl the top apps by examining the apps listed on the top chart for each category. The second way is to start on the main page, and then follow each link that is either another app, a developer, or a category. This provides a much more thorough crawl of apps. Having both of these methods as options allows us to perform quicker crawls to keep data on multiple versions of the more popular apps and to have a more thorough crawl. Both of these methods utilize multithreading to take advantage of concurrency in an I/O-bound task.

As mentioned previously, we are not the first to do a large-scale crawl of the Android app marketplace. Due to time constraints and the rate-limiting enforced by Google on calls to their API endpoint, we were not able to complete a full crawl with the second crawling method discussed in the previous paragraph. However, we still were able to collect data for over 1.9 million apps over the past 10 years, 364,821 of which currently exist on Google Play. This covers 13.3% of the estimated 2.7 million apps that are currently listed on the store.

*4.1.2 Scraper.* The scraper takes the package names retrieved by the crawler and uses an unofficial Python API for Google Play to retrieve a thorough set of metadata for each app [20]. To avoid being rate-limited, we cycle through a list of ten accounts, generating a new token for the selected account when an error message indicating throttling is received. In addition, we also catch any 404 HTTP errors and use that to track if an app has been removed from Google Play. For similar reasons as the crawler, the scraper uses multithreading for concurrency.

*4.1.3 Downloader.* The downloader takes the package names of apps that have not been marked as removed and downloads the APK through a Python library called `gplaycli, gplaycligit`. We extended this library to use the same login and token generation mechanism as the unofficial Python Google Play API for convenience. Due to the I/O-intensive nature of this module, it uses multithreading to achieve concurrency as well.

*4.1.4 Decompiler.* The decompiler uses `apktool` [33] and simply makes a call to the necessary shell command. The decompiled contents are then compressed and stored on our NAS. However, this step is only run in the pipeline for apps that were found in the top charts of a given category. Otherwise, decompiling an APK is done on an as-needed basis. Given that this is a computationally-intensive workload, this step is parallelized across many processes to take advantage of the high number of cores allotted to the VM.

*4.1.5 Analyzer.* The analyzer conducts our static analysis on the downloaded APKs, the core of which is comprised of analyses of the third-party libraries, permissions, and URLs or links used by each app. The static analysis portion also utilizes the plugin mechanism to analyze the different types of identifiers used by an app, and whether or not it scans for other installed apps on the device. Many of these analyses are enabled by AndroGuard [11], which is a tool for reverse engineering and statically analyzing Android apps. Similar to the decompiler, this step is parallelized across many processes due to its computationally intensive nature.

## 4.2 API Layer

The API component was a prototype of how the data could be shared with the broader research community to facilitate easier research on Android apps. To demonstrate its usability, the analyses presented in the remainder of this paper were all done through data retrieved from this API layer.

## 5 ASPECTS OF APP PRIVACY EXAMINED

In this section, we discuss the aspects of privacy that we chose to consider and the rationale behind our decision. These aspects include the adoption of HTTPS by apps, whether apps scan the device for other installed apps, the use of permissions to request privacy-sensitive data, and the use of various types of unique identifiers. For each aspect, we expand on what it entails, why it is important to privacy, and any related previous work.

As mentioned earlier, there were several key reasons why we opted to focus on these aspects of privacy. The first is that most of these factors are common across many apps; for example, permissions and unique identifiers are both often used for providing desired functionality, while HTTPS should always be the default protocol for network communication. Second, our analysis methods allowed us to determine whether it was the app's internal behavior or a third-party library that was responsible, which was necessary to investigate who was accountable for changes in privacy behavior. Finally, these factors were easy to analyze at scale because the analysis methods were not extremely computationally intensive and apps were able to be analyzed in parallel.

### 5.1 HTTPS Adoption

HTTPS protects the privacy of a user's data by adding a layer of security to network communication over regular HTTP. This prevents any intruders or other parties from surveilling the connection and snooping potentially sensitive data, for example through man-in-the-middle (MITM) attacks. There has long been a push to adopt HTTPS, either with SSL or TLS, and its usage is especially important for mobile device apps because mobile devices frequently use unsecured networks, such as public networks or hotspots [13]. Google's Android Developers guide on networking security recommends that HTTPS be used over HTTP anywhere it is supported by the server [13].

Previous work studying the adoption of HTTPS in Android apps discovered that HTTPS adoption was very slow, even when the data being communicated over the network contained PII [29]. Other works also found that apps that adopted HTTPS were still susceptible to MITM attacks [17] and/or had vulnerabilities due to misconfigurations or weak cipher suites [28]. Compared to these past works, we provide a larger scale analysis as we examine HTTPS adoption over nearly 20,000 apps, as opposed to only hundreds to thousands. Furthermore, we analyze how the adoption of HTTPS by the internal code of apps differs from the third-party libraries used.

### 5.2 Scanning for Other Installed Apps

The Android platform provides a way to retrieve a list of all other apps installed on the user's device through two specific API calls, `getInstalledPackages` and `getInstalledApplications`. This means that apps can retrieve a list of all other apps installed, and these calls are not protected by a permission so users do not have to consent to this behavior. This type of behavior is a problem for user privacy because this information can be used to fingerprint a device and therefore a user. It can also leak potentially sensitive information a user might not want to share, e.g. a pregnancy monitoring app or an Alcoholics Anonymous app. In addition, a user does not have to explicitly provide consent for an app to scan for other apps installed on the device. There is also no explicit policy from Google about the usage of these APIs and the scanning of installed apps on a device.

Pham et al. analyzed the usage of these API calls in the context of them being used to fingerprint mobile devices [26]. They found that over half of the 2,917 apps they examined did use these APIs but only a quarter of those did so due to the internal code of the apps [26]. Similar to our analyses of HTTPS adoption, our work is novel in that it analyzes these APIs at a much larger scale, examines longitudinal changes, and observes the differences in the usage of these APIs between the internal code of apps and third-party libraries.

## 5.3 Sensitive Permissions

Permissions are a critical component of how the Android operating system protects the user by being the gateway for accessing the user's data and/or the device's features. Some of these protect privacy-sensitive resources and we refer to these as sensitive permissions. The list of sensitive permissions that we investigate in this paper were taken from two sources. The first is Google's public list of permissions they consider as dangerous, which are specific permissions that request data or resources involving the user's private information or potentially affect stored data or the operation of other apps [12]. The second is the Pew Research Center's list of permissions that access user information [4]. The full list of sensitive permissions we consider is shown in Table 8 in the appendix.

Sensitive permissions can pose an issue with privacy because, once granted, the app is able to access the requested data, which potentially contains PII. In addition, the app, or a third-party library used by the app, could use the permission in a way that is unexpected or unrelated to its functionality [22]. Many users also do not thoroughly read the description or examine the permissions used when installing an app [22].

There has been extensive past work studying how Android apps use permissions. Some of this work has focused on whether the permissions used by an app are necessary based on the app's purpose or description [18, 25]. Other work has looked at the contexts in which the APIs for permissions are used to determine if an app is actually malware [38]. There have also been previous works examining the change in usage of sensitive permissions over time by third-party libraries or pre-installed apps [10, 36]. Our work offers a different perspective by doing a larger scale longitudinal analysis, and by directly comparing the usage of sensitive permissions by an app's internal code against the third-party libraries used.

## 5.4 Types of Identifiers

There are various reasons for an app to uniquely identify the user and/or the device, such as personalization, user profiling, and ad tracking. There are three main types of unique identifiers, which are advertising IDs, Android IDs, and hardware identifiers. An advertising ID is generated by the `AdvertisingIdClient` and is user-resettable, and not connected to PII. As of Android 8.0, or Android Oreo, the Android ID is unique to each combination of app, user and device; however, prior to this it was a constant value that was persistent across the lifetime of the device. On the other hand, hardware identifiers are based on information about the device's hardware, such as its MAC address or IMEI number. Because it is based on a device's actual hardware, it is completely persistent.

Google has published a list of best practices regarding the use of unique identifiers, and has enforced the usage of only an advertising ID for user profiling and ad tracking since 2014 [9, 15, 19]. Although these different types of unique identifiers do not leak any PII, they still pose an issue to privacy as they are persistent across a device's lifetime and can be linked to the specific user. This takes away the user's ability to reset their tracking profile and control other options regarding advertising and tracking. On the other hand, even though the advertising ID is user-resettable, an app could still bridge these resets by connecting previous and current IDs, or even choose to ignore a user's advertising settings.

Previous work has found that developers were unaware of the privacy ramifications of the identifiers they used and of better alternatives, especially for hardware-based identifiers [21]. In addition, Ren et al. showed that Google's enforcement of the usage of advertising IDs led to apps using advertising IDs instead of other types of identifiers, but also that third-party domains receive multiple types of unique identifiers to continue to identify users [29]. Our work provides new data at a large scale on how the usage of unique identifiers has changed over time, and how this differs between the internal code of apps and third-party libraries.

# 6 METHODS OF ANALYSIS

This section discusses how we gathered and analyzed the data for each aspect of privacy we chose to focus on.

## 6.1 AndroGuard

One of the key tools we used throughout the various analyses was AndroGuard [11], which, as mentioned before, is a tool for reverse engineering and statically analyzing Android apps [11]. AndroGuard was released many years ago but it still maintains its accuracy due to constant updates from open-source contributors. Although AndroGuard is very versatile, we heavily utilized one feature in particular. This feature was AndroGuard's concept of crossreferencing (XREF), which was formerly known as taint analysis. This involves tracing important resources through the bytecode of a decompiled APK to discover which ones are actually referenced by the code and where. This enables analyses beyond just what is declared in the manifest, such as analyses of which portions of the code or which libraries are responsible for what behavior. We use XREFs to find the usage of specific strings, Java methods, and Java classes. For example, we used the XREFs generated for strings to determine links used in the app, and the XREFs for methods and classes to determine the usage of specific APIs or system functions.

In addition to XREFs, we also used AndroGuard to determine whether the string, method, or class in question was used by the app internally or by an external third-party library. To do so, we used AndroGuard to parse out the path to where the resource was used in the app, then compared that against the main path of the app. The main path contains the package name of the app itself. Any paths that had a match with the main path was considered as internal use by the app's code, and any others were considered as use by a third-party library.

## 6.2 Methods

To analyze how HTTPS adoption has changed in Android apps, we analyzed the XREF strings of an app and parsed the strings to determine which ones were valid URLs or links. Then, we checked which strings began with `http://` and which with `https://`. This gave us an approximation of how widely HTTPS was being adopted. We note that this did not account for the protocol of any potential network communication done through third-party library methods, such as OkHttp [30]. However, it did account for any communication using Android system APIs as they require the full URL to be specified, which includes the protocol identifier [14]. We leave the detection of HTTPS adoption from third-party library methods to future work. To remove the effect of an app reusing the same URL multiple times and reduce the effect of any URLs that might just be dummies, we only considered unique combinations of URLs and external library name for each app. If the URL was used by the app's internal code and not a third-party library, then the library name was null. Furthermore, we broke down an app's usage of HTTPS by both the number of domains and the number of URLs.

We analyzed the occurrence of apps scanning for other installed apps by using XREF methods and classes. From the Android developers documentation, we found that the `getInstalledPackages` and `getInstalledApplications` methods were the APIs that provided a list of other apps installed. As such, we looked for apps that used either of these APIs.

We analyzed the usage of sensitive permissions in two ways, which from here on forward will be referred to as *manifest usage* and *tainted usage*. One way we could have analyzed *manifest usage* would be to decompile the app's APK and parse out the sensitive permissions listed in the `AndroidManifest.xml` file. However, the metadata we scraped for each app also contained this information so we instead conducted our analysis from that data. *Tainted usage* represented the permissions actually used based on the methods and classes XREFs generated. From these methods and classes, AndroGuard is able to use an existing mapping of Android APIs to permissions to determine which permissions are actually referenced by the bytecode. The tainted usage of sensitive permissions is the subset of these referenced permissions that are also considered as sensitive permissions. Since the manifest usage method was actually determined from the app metadata and did not require the APK, our analyses using the manifest usage uses a larger set of apps than the rest of the factors of privacy.

To analyze the use of unique identifiers, we determined the classes and methods needed to retrieve each of them through the Android developers documentation. We then compared the XREF methods and classes of an app with the desired ones to determine the types of identifiers used for each app. In addition, we also differentiated between current and older methods for retrieving the various identifiers, where applicable. For example, while there are different recommended methods for retrieving the device identity number for different API versions, there is one method for retrieving the MAC address that has not been entirely deprecated.

## 7 CHANGES IN PRIVACY OVER TIME

The first question we consider here is how app behaviors have changed over time with respect to these specific aspects of privacy. From the apps in our data set, we found that privacy-related behavior has generally seen an improvement with time as apps mature and receive more updates. Specifically, we found a decrease in the number of apps that scanned for other installed apps and a decrease in the number of types of unique identifiers being used.

### 7.1 Approach

For the rest of the paper, we define the concept of an app's *maturity* as the amount of time between the oldest and newest versions of an app. We focused on apps that we have multiple versions of data for, compared the oldest and newest version of these, and bucketed them by their maturity expressed in terms of months. Specifically, we only analyzed the 18,446 updated apps that had an update after August 1st, 2018. We chose this date to be roughly a year prior to the writing of this paper, and considered any other updated apps that did not meet this criteria to be an app that did not survive. For each of the aspects of privacy, we plotted the data against app maturity and fit each of them with a linear regression to assess trends in changes.

### 7.2 Little Change in Use of HTTPS by Apps

Figure 4 shows the average adoption of HTTPS for domains and for unique URLs as the maturity of apps vary. The data covered 18,446 apps that had been updated. Although the trend line shown by the linear regression demonstrated a slight decrease for both, there was not a strong fit for either. The percentage of domains for which HTTPS was adopted for ranged from 46.8% to 68.5%, while the percentage of unique URLs ranged from 53.5% to 79.3%. Therefore, neither aspect of HTTPS adoption showed a clear change with increasing app maturity. However, in the current state of these apps, the majority of network communication we found did utilize HTTPS, with an average of 72.5% of the links and 61.9% of the domains using it.

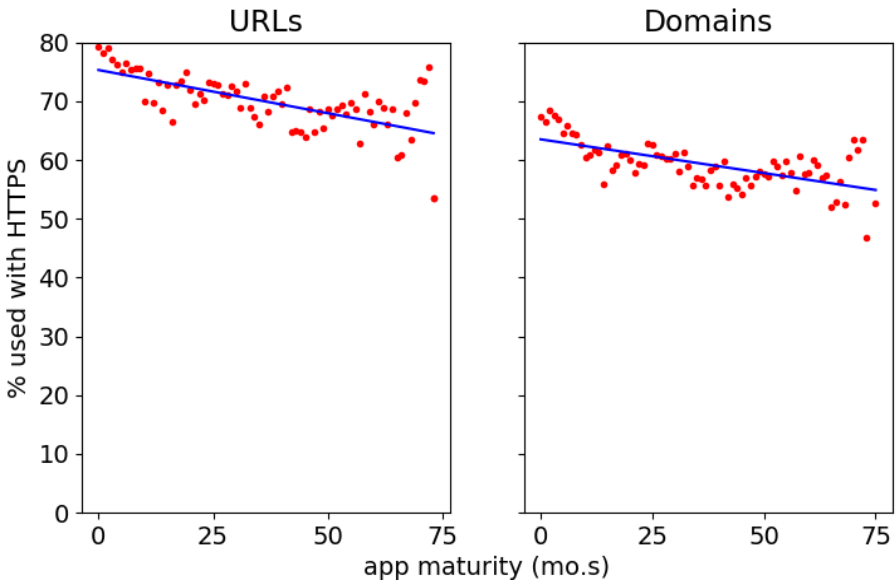

Fig. 4. Across 18,446 updated apps, the average adoption of HTTPS on both a domain basis and URL basis showed no clear change with increasing maturity, as apps across the board displayed usage percentages in the same range of values. App maturity is calculated as the amount of time between the oldest and newest versions of an app, in months.

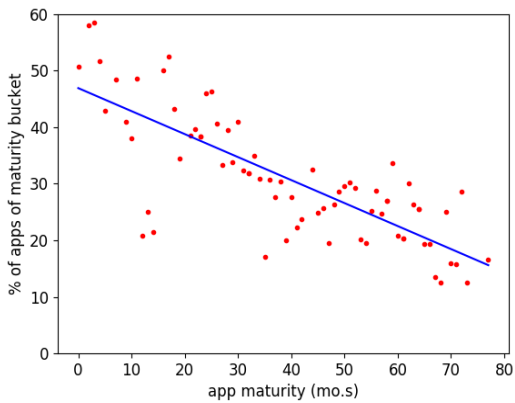

Fig. 5. The decrease in occurrence of scanning for installed packages across the 18,446 updated apps showed strong correlation with increasing app maturity.

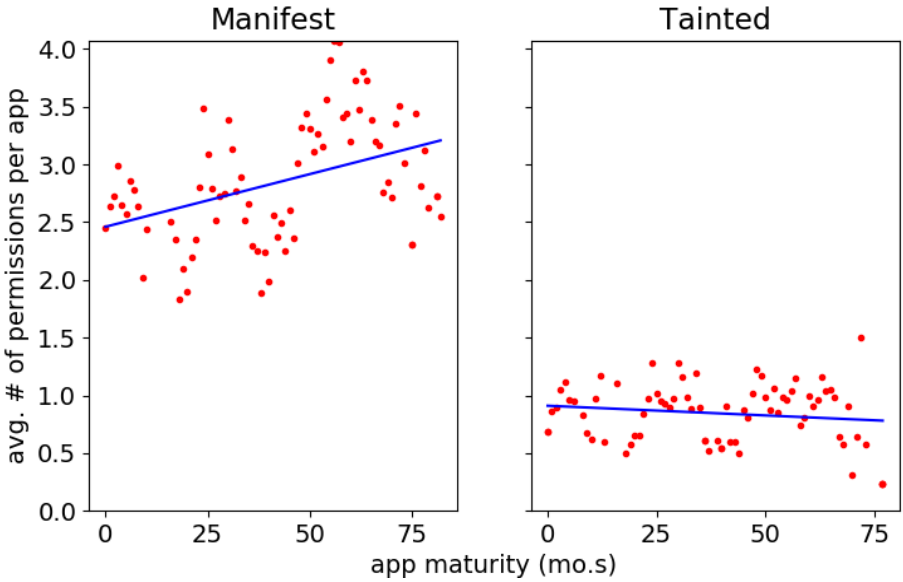

Fig. 6. The average number of sensitive permissions used with increasing maturity had no clear trend among the 18,446 updated apps for tainted usage. However, the data demonstrated a slight increase for manifest usage when looking at a broader set of 65,858 apps.

### 7.3 Decrease in Scanning for Other Installed Apps

In contrast, a much clearer trend presented itself when examining how frequently apps use the APIs for scanning for other installed packages on the device, as shown in Figure 5. The trend line showed a clear decrease among the 18,446 updated apps that scan for other installed packages as app maturity increased. Specifically, we can expect a decrease of 0.41% in the number of apps that exhibit this behavior for each additional month of time an app has been on Google Play. Although there were still a small number of more mature apps that conducted this behavior, the data demonstrated that apps' behaviors with respect to privacy in this aspect does improve with time. Furthermore, even though currently 40.8% of these apps do scan for other installed apps, this could potentially be attributed to third-party libraries as opposed to the apps themselves, which we discuss in a later section.

### 7.4 Increase in Use of Sensitive Permissions

As discussed in section 5.3 of this paper, we conducted an analysis of sensitive permission usage from the perspective of manifest usage and tainted usage. Manifest usage is based on the permissions declared in the manifest file of an app while tainted usage is based on the XREFs generated by AndroGuard. Whereas we conducted the tainted usage analysis over 18,446 updated apps, we were able to conduct the manifest usage analysis over a broader set of 65,858 apps since this information was also available in the scraped metadata for each app. Since there are many different sensitive permissions to consider, we focused only on the number of different sensitive permissions used by an app, and the individual sensitive permissions that demonstrated strong and/or interesting trends.

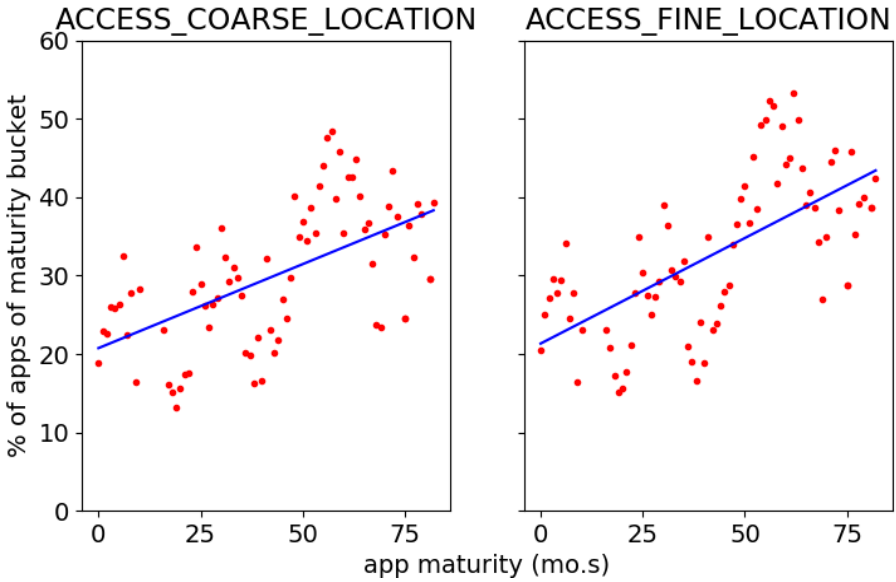

Fig. 7. For the manifest usage of sensitive permissions, across 65,858 updated apps, the ACCESS_COARSE_LOCATION and ACCESS_FINE_LOCATION permissions showed an increase in the number of apps that used them with increasing maturity.

Figure 6 shows how the number of sensitive permissions used varied with app maturity. Although there was no clear trend of change for tainted usage, the manifest usage of these permissions demonstrated a slight increase with increasing app maturity. There was also a significant difference in the usage of permissions when comparing the manifest and the tainted usages. This is because the manifest usage is based off of the permissions declared by a developer in `AndroidManifest.xml` but developers may not necessarily use the corresponding API for every declared permission. The tainted usage analyzed permissions based on the actual calls made to the APIs and therefore shows a lower usage. This could also potentially be due to incomplete mappings between APIs and permissions as there is no published mapping from Google.

Similar to the number of sensitive permissions used, there were no specific permissions that displayed interesting trends when considering tainted usage. However, two specific permissions that showed interesting trends with manifest usage were the ones for getting a user's location data, ACCESS_COARSE_LOCATION and ACCESS_FINE_LOCATION, as shown by Figure 7. Both of these permissions showed an increase in the percentage of apps that use them as app maturity increases. The trends showed an average increase of 0.21% of apps per month that used ACCESS_COARSE_LOCATION and an average increase of 0.27% per month that used ACCESS_FINE_LOCATION. The other sensitive permissions did not show any interesting trends.

For the most current versions of these apps, we found each app uses on average 3 permissions when looking at the manifest and 1 permission when looking at the corresponding API calls. Table 2 also shows the top 5 most used sensitive permissions currently.

| Permission | % of apps that currently use |
|---|---|
| READ_EXTERNAL_STORAGE | 67.3 |
| WRITE_EXTERNAL_STORAGE | 65.8 |
| ACCESS_FINE_LOCATION | 34.6 |
| READ_PHONE_STATE | 34.1 |
| ACCESS_COARSE_LOCATION | 30.8 |

Table 2. These are the 5 most commonly used sensitive permissions declared through the manifest among the current versions of the updated apps.

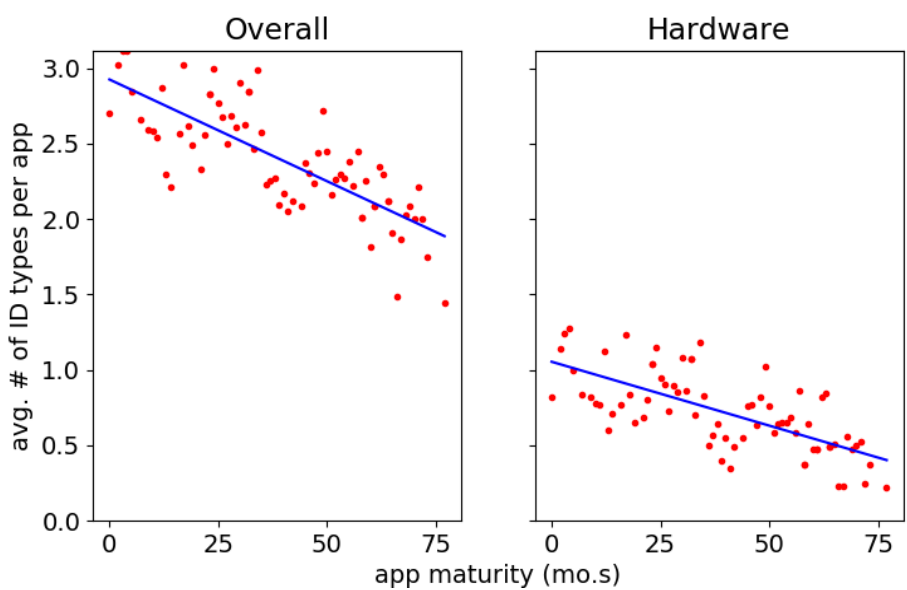

Fig. 8. Among the 18,446 updated apps, the average number of unique identifier types used has decreased vs increasing maturity. This was true for all types of identifiers, as well as just the hardware-based ones.

## 7.5 Decrease in Use of Unique Identifiers

Similar to our analysis of sensitive permissions usage, we chose to focus our discussion on the usage of unique identifiers on the number of different types used and specific types with notable trends. Figure 8 shows how the number of types of identifiers used, as well as the number of types of hardware-based ones used, has varied with app maturity for 18,446 updated apps. The data showed that both of these have decreased with increasing app maturity, which means that privacy-related behavior has improved in this aspect because apps are using less ways to track and profile users. This is especially true for the decrease in usage of hardware-based identifiers as the only identifier that should be used for most apps is the advertising ID.

Certain specific types of unique identifiers also showed a decrease in usage with increasing maturity, as shown by Figure 9. The types that did so were advertising IDs, Android IDs, and device MAC addresses. On average, 0.23% less apps used the Android ID, and 0.38% less apps used the MAC address of a device, per month of app maturity. The decreases in the usage of these two unique identifiers further support the fact that privacy-related behavior improves with app maturity in this aspect.

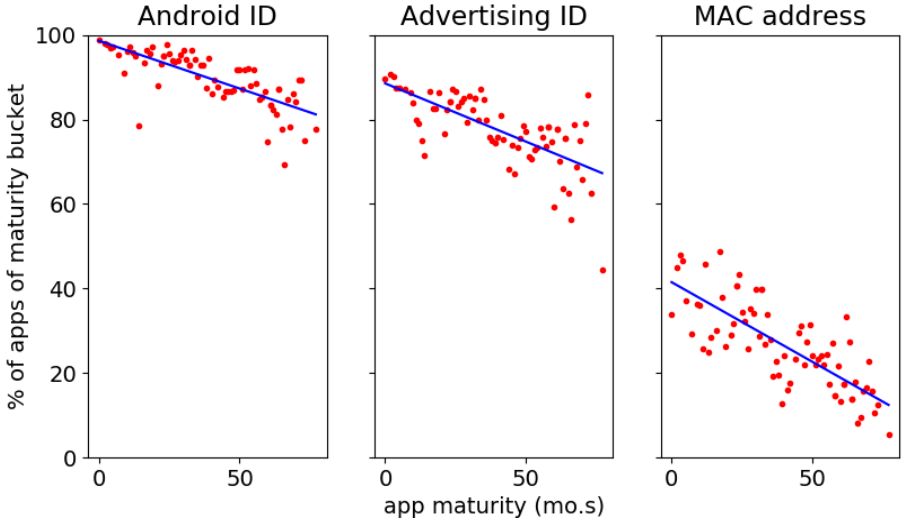

Fig. 9. The advertising ID, Android ID, and MAC address unique identifiers exhibited a decrease in the number of apps that use them with increasing maturity.

When looking at the current state, we observed that each app uses an average of 2.7 types of unique identifiers and 1 type of hardware-based unique identifier. We also found that the same 3 specific types of identifiers discussed above (advertising ID, Android ID, MAC address) are also currently the 3 most used types. 93.1% of the apps currently use an Android ID, 82.0% use an advertising ID, and 34.6% use the device's MAC address.

## 8 ACCOUNTING FOR CHANGE: APPS OR LIBRARIES

In previous work, third-party libraries have been cited as a major source of issues with poor privacy-related behavior for apps, as opposed to the app's own internal code [2, 6, 8, 10, 32]. The usage of these third-party libraries is also immensely popular because of their ability to conveniently enhance an app's features [6]. A previous study found that over 51% of 1,100 free apps used some type of ad or analytics library [16], while another estimated that at least one ad library was installed in hundreds of thousands of free apps [10]. The combination of these two factors are what makes the analysis of third-party libraries and their role in privacy issues very important. In this section, we investigate how third-party library usage affects the specific aspects of privacy we are examining. Our analysis indicated that third-party libraries still accounted for many of the privacy issues we observe in apps, and that there has been little improvement in good privacy practices over time.

### 8.1 Approach

Since we were interested in both how the impact of the responsible parties of privacy issues has changed over time and how it can be accounted for in the current state of Android apps, we gathered data on both older apps and recent apps. We considered apps uploaded on or before June 1st, 2017 to be older and apps uploaded on or after January 1st, 2019 to be recent. This resulted in a data set of 60,060 apps, with 28,638 older apps and 31,422 recent apps. For this section, we looked at all apps in general, not just ones that have been updated. This allowed us to pick different date thresholds to have a larger difference in time between the old and the recent sets of apps.

| Top 5 Most Common Libraries | |
|---|---|
| **Library** | **% of Apps That Use** |
| admob | 92.0% |
| GoogleAnalytics | 60.0% |
| firebase | 47.0% |
| facebook | 42.3% |
| squareup | 35.4% |
| **Top 5 Most Privacy Issue Libraries** | |
| **Library** | **% of Apps That Use** |
| facebook | 42.3% |
| squareup | 35.4% |
| amazon | 20.2% |
| github | 17.2% |
| unity | 5.45% |

Table 3. The percentage of apps we found these libraries used in ranged all the way from around 5% to around 92%. The overlap of libraries between the most common and the ones with the worst privacy behavior is important to note as it shows that the issues of a single library can affect many apps.

For each app, we examined its behavior with respect to the same aspects of privacy as previously discussed and whether the behavior was caused by the app's internal code, a third-party library, or both. Our analyses focused on the proportion of apps that exhibit a given privacy issue for each party, as opposed to comparing the issues by unique apps and the issues by third-party libraries. To identify the responsible party of the behavior, we again used the XREFs generated by AndroGuard and compared the path to where the behavior was observed against the main path, which contains the package name of the app itself. We also assumed that no obfuscation was used for the apps. In addition to the comparison between app internals and third-party libraries, we also examined the current state of privacy behavior of the top 5 most widely used libraries and the 5 with the most privacy issues. Table 3 shows how widely used each of these libraries were.

## 8.2 Libraries Drag Down HTTPS Adoption Efforts

Figure 10 shows the average adoption of HTTPS for domains and for unique URLs overall, as well as for apps internal code and third-party libraries. As we can see, overall adoption of HTTPS was not as widespread as it should have been. HTTPS usage for domains has only increased from 35.3% to 49.2% of domains, and from 64.3% to 71.3% for unique URLs. However, the data showed that in the current state of Android apps, libraries have adopted HTTPS for fewer domains than apps' themselves internally. Although the opposite was true in the past, libraries are now the ones more responsible for low adoption. Figure 10 also shows that libraries have consistently adopted HTTPS for fewer URLs than apps' themselves. Combined, we can see that libraries have demonstrated both a less widespread and a lower rate of adoption of HTTPS.

Given this conclusion, we also investigated the adoption of HTTPS of some specific libraries in particular, which is shown in Table 4. We found that all of the top 5 most commonly used libraries at this current point in time have adopted HTTPS for less than half of their network communication, with it being as low as less than a quarter for some. When looking at the 5 libraries that caused the most privacy issues, we found that 2 did actually utilize HTTPS for a majority of their network flows, but the other 3 only used it for at most 36.2% of their flows.

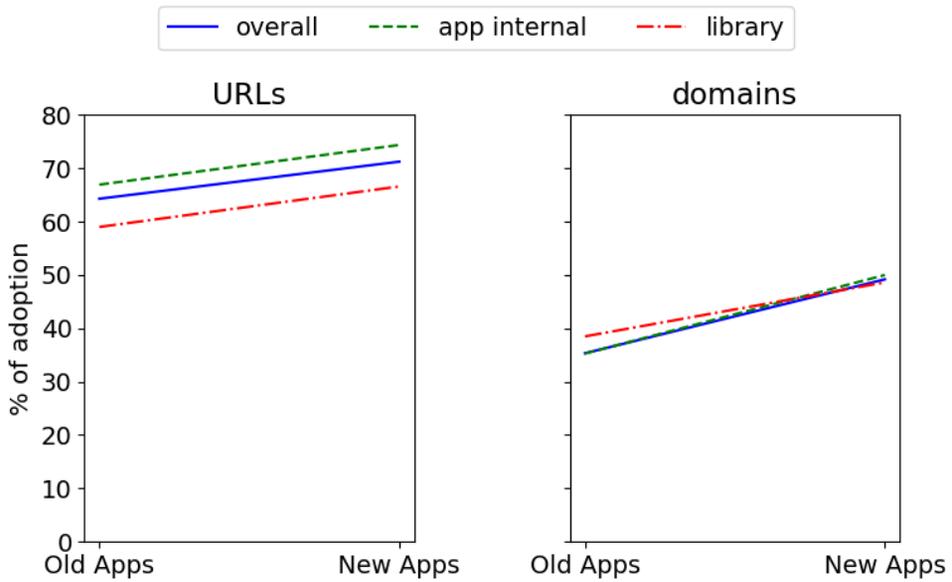

Fig. 10. Adoption of HTTPS is not as widespread as it should be across the 60,060 apps analyzed. In terms of which party contributes more to this issue, third-party libraries are the ones who have adopted HTTPS less on both a domain basis and URL basis.

| Top 5 Most Commonly Used Libraries | | |
|---|---|---|
| **Library** | **% of HTTPS Links** | **% of HTTPS Domains** |
| admob | 43.0 | 37.8 |
| GoogleAnalytics | 12.1 | 28.6 |
| firebase | 4.9 | 29.8 |
| facebook | 27.1 | 36.2 |
| squareup | 25.4 | 11.5 |
| **Top 5 Libraries with Most Privacy Issues** | | |
| **Library** | **% of HTTPS Links** | **% of HTTPS Domains** |
| facebook | 27.1 | 36.2 |
| squareup | 25.4 | 11.5 |
| amazon | 8.4 | 20.5 |
| github | 62.9 | 55.8 |
| unity | 95.4 | 50.0 |

Table 4. This table shows current level of adoption of HTTPS for the top 5 most commonly used libraries and the 5 libraries with the most privacy issues. This shows that libraries do pose an issue with privacy because of the low level of adoption, even in the most commonly used libraries.

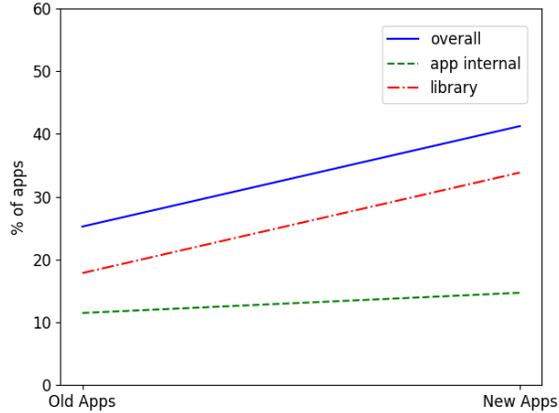

Fig. 11. A concerningly large number of apps out of the 60,060 apps scan for packages installed on a device, and the number is growing. Third-party libraries are more likely to be the source of this behavior and have been the ones driving the increase in it.

### 8.3 Libraries are Primary Cause of Scanning for Installed Apps

There are only a few types of apps that would justifiably need to scan a user's device for the other apps installed, such as notification or storage managers [37]. Therefore, the large number of apps that have exhibited this behavior in the past and in the present, as shown by Figure 11, is worrying for user privacy. The data indicated that this behavior increased from being observed in 25.2% of the 60,060 apps to 41.3%. In addition to leading to a slower adoption of HTTPS in apps, third-party libraries have also been the driver for the increase in apps that scan for other installed apps on the user's device. While the number of apps where the app's own internal functionality was responsible for this behavior only grew from 11.5% to 14.7%, the number of apps where a third-party library was responsible grew from 17.8% to 33.9%. The increase due to third-party libraries is clearly much steeper, and therefore they have been the main driver of the growth of this behavior.

Table 5 shows how many of the specific libraries investigated retrieve for other installed apps. We found that all but one of those libraries did use at least one of those APIs. While this was expected for the libraries that had the most privacy issues overall, this is a concerning observation regarding the most commonly used libraries. However, it further supports our conclusion that libraries are what drives the growth in the usage of these APIs to scan the installed apps on a device.

### 8.4 Unnecessary Sensitive Permissions Used by Libraries More

Looking at sensitive permissions from both manifest usage and tainted usage perspectives, there has been an increase in the number used when comparing older apps to their more recent versions. This is shown in Figures 12 and 13. As mentioned before, we were able to examine the manifest usage across a broader set of apps than the tainted usage because a downloaded APK was not needed to analyze the manifest usage of sensitive permissions. In this instance, we conducted the tainted usage analysis over 60,060 apps and the manifest usage analysis over 305,148 apps. When looking at the manifest, 64.4% of the older apps used 1 or more sensitive permissions, but this increased to 80.0% for the recent apps. This trend was also true no matter how the usage of sensitive permissions was accounted for, whether it was the apps' internal code, third-party libraries, or both. For tainted usage, the percentage of apps that used one or more sensitive permissions increased from 41.3% to 50.0% overall, from 35.4% to 44.8% for internal app code, and from 25.3% to 34.5% for third-party libraries.

| Top 5 Most Commonly Used Libraries | |
|---|---|
| **Library** | **Scans for Installed Apps?** |
| admob | Yes |
| GoogleAnalytics | No |
| firebase | Yes |
| facebook | Yes |
| squareup | Yes |
| **Top 5 Most Privacy Issue Libraries** | |
| **Library** | **Scans for Installed Apps?** |
| facebook | Yes |
| squareup | Yes |
| amazon | Yes |
| github | Yes |
| unity | Yes |

Table 5. This table shows the usage of the APIs for listing installed apps by the top 5 most commonly used libraries and the 5 libraries with the most privacy issues. Almost all of the libraries do use at least one of these APIs, posing a serious issue with privacy.

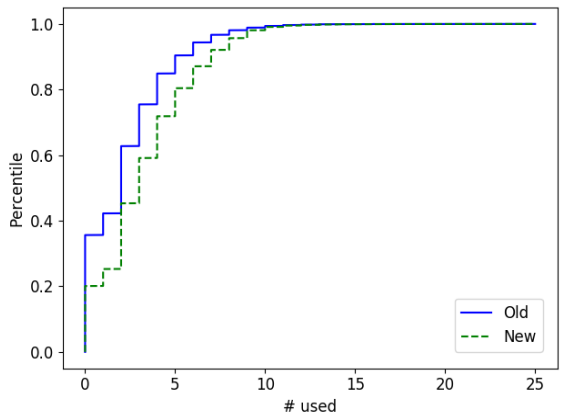

Fig. 12. This CDF shows that the number of sensitive permissions used based on manifest usage has increased between the older and the recent sets of apps across 350,148 apps.

Although apps internally and third-party libraries show the same trends in terms of the number of sensitive permissions used, we observed differences in the specific ones that were used. This is shown in the graphs in Figure 14. Apps internally used more sensitive permissions to retrieve information around location and device accounts. On the other hand, libraries exhibited a higher usage of the permission to gather data about the state of the device's phone operation, such as phone number and cellular network information.

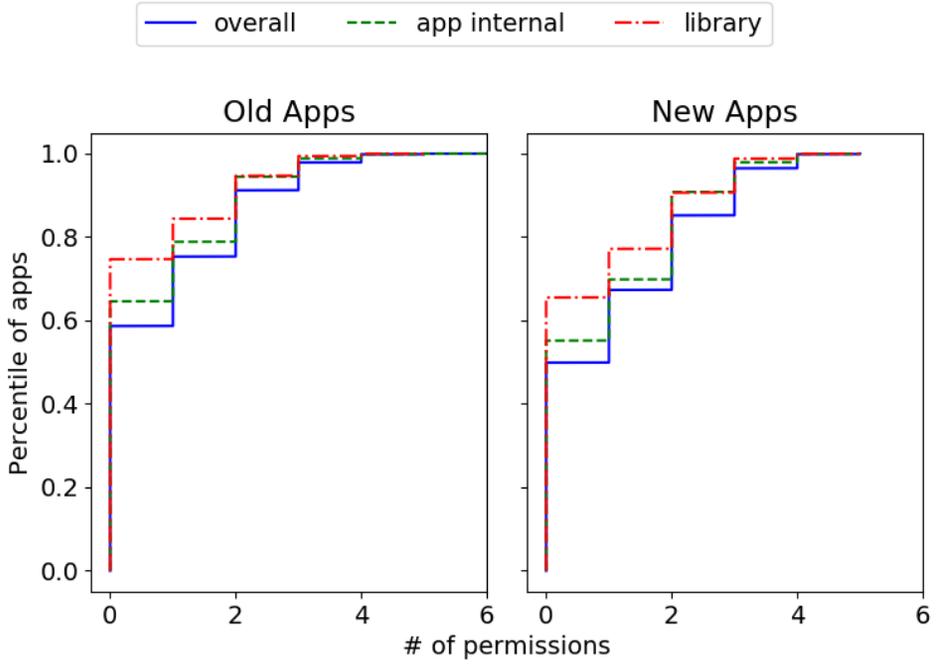

Fig. 13. Across the 60,060 apps, the number of sensitive permissions used with tainted usage has increased on an overall basis, and has increased for each responsible party as well, as shown by this CDF.

| Top 5 Most Commonly Used Libraries | |
|---|---|
| **Library** | **# of Sensitive Permissions Used** |
| admob | 2 |
| GoogleAnalytics | 0 |
| firebase | 2 |
| facebook | 4 |
| squareup | 4 |
| **Top 5 Libraries with Most Privacy Issues** | |
| **Library** | **# of Sensitive Permissions Used** |
| facebook | 4 |
| squareup | 4 |
| amazon | 5 |
| github | 4 |
| unity | 3 |

Table 6. This table shows how many sensitive permissions are used via tainted usage by the top 5 most commonly used libraries and the 5 libraries with the most privacy issues. We found that these common libraries still use a large number of sensitive permissions, which poses a privacy risk to the end users.

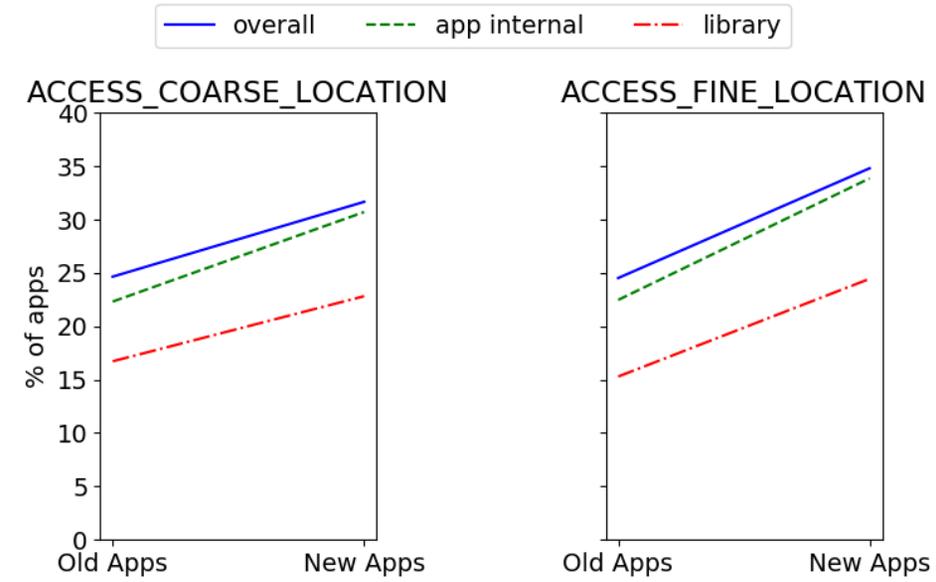

(a) Usage of ACCESS_COARSE_LOCATION and ACCESS_FINE_LOCATION.

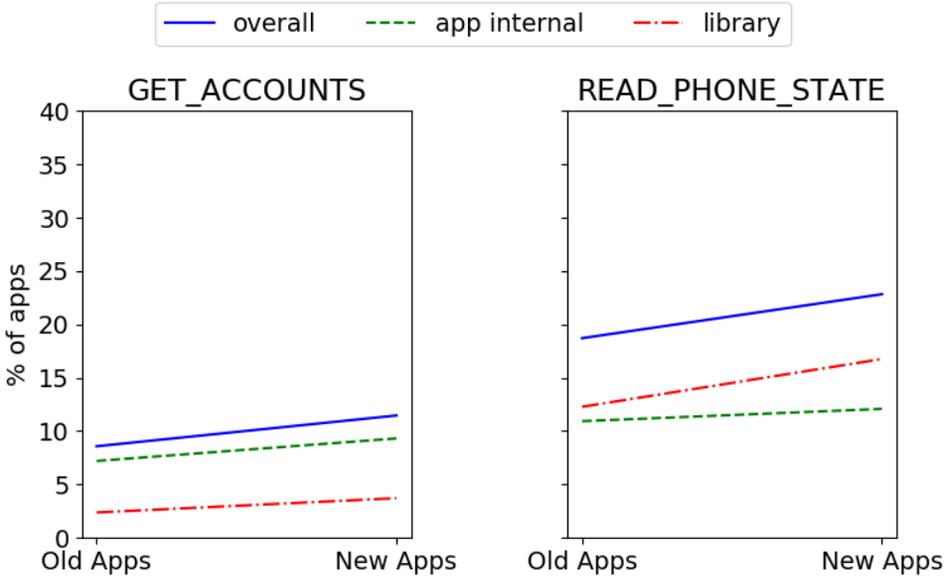

(b) Usage of GET_ACCOUNTS and READ_PHONE_STATE.

Fig. 14. From examining 60,060 apps, we found that the specific sensitive permissions used more frequently by apps' internally and by third-party libraries were different.

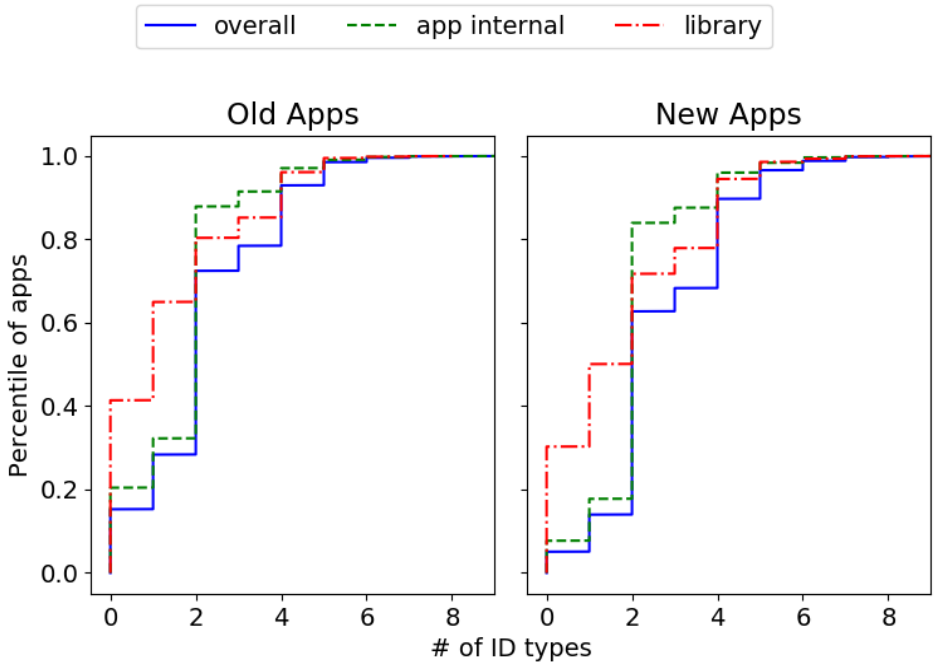

Fig. 15. When examining the number of unique identifier types with respect to the usage by each accountable party, the CDF demonstrated that apps' internally were most likely to use 2, while third-party libraries were more likely to use less than and more than 2 types.

Table 6 shows the number of sensitive permissions used by specific libraries we investigated through the tainted usage analysis method. As expected, the libraries with the most privacy issues used sensitive permissions more egregiously; the usage of 3 or more sensitive permissions puts each of these libraries into the 90th percentile of Figure 13. One point of interest in Table 6 is that two of the most commonly used libraries overlap with the set of libraries with the most privacy issues. This shows that many apps that use either of these libraries are already requesting for a large number of permissions from the user, posing a privacy risk to users.

### 8.5 Libraries Use Many Unique Identifiers Frequently

Figure 15 shows how the number of different unique identifiers has changed from the set of older apps to the recent ones, due to apps internally and third-party libraries. Both sets showed similar trends with respect to the comparison between app internal usage and third-party library usage. Out of the 60,060 apps, apps' internally were much more likely to use 2 types of identifiers, with 55.6% of older apps and 66.2% of recent apps doing so. However, third-party libraries were more likely to use any other number of types, with 65.0% of older apps and 50.0% of recent apps using less than 2, and 19.6% of older apps and 28.2% of recent apps using more than 2. From these results, we also observed that usage of identifiers has increased overall between these sets of apps.

We also analyzed how apps used hardware-based unique identifiers compared to other identifiers. As discussed earlier in section 5.4, these identifiers are more privacy invasive because their values are unique to the user's device, not a given combination of app and device, and cannot be reset by the user. Figure 16 shows how the number of hardware-based identifiers used by apps internally and by libraries has changed. We can see that third-party libraries were much more likely to use hardware-based identifiers. 13.8% and 21.7% of older apps used more than 1 identifier due to app internals and third-party libraries, respectively. Similarly, 17.0% and 30.3% of newer apps used more than 1 identifier due to app internals and third-party libraries, respectively.

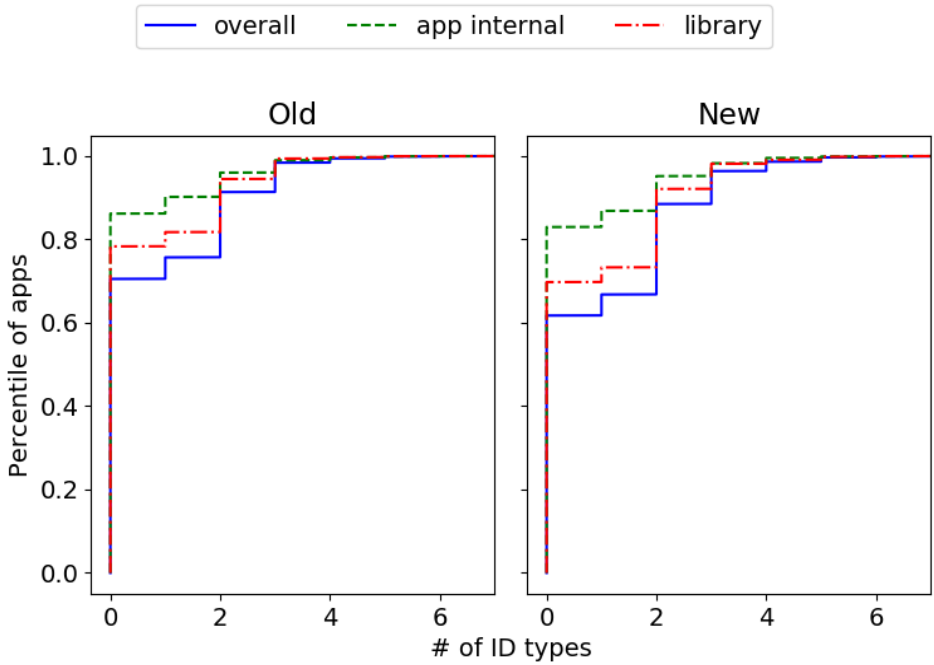

Fig. 16. Third-party libraries are more likely to be the source of the usage of hardware-based unique identifiers than the internal code of apps, as shown by this CDF.

This trend was also clear when looking at the usage of specific hardware-based identifiers, as shown in Figure 17, and comparing them against the usage of ones that are not hardware-based, as shown in Figure 17b. Third-party libraries exhibited higher usage of device identity numbers and MAC addresses than apps' internal code, as well as a steeper increase in their usage of these identifiers. On the other hand, libraries only used advertising IDs and Android IDs at most half as often as apps did internally.

Table 7 shows the number of types of unique identifiers and of only hardware-based identifiers used by the specific libraries we investigated. Similar to the usage of sensitive permissions, libraries with the most privacy issues utilized a large number of different types of unique identifiers; some even used almost all of the ones documented by Google. Libraries that used 6 or more types of identifiers, or 2 or more hardware-based identifiers, were in the 99th percentile of Figures 15 and 16, respectively. Again, the overlap between the most commonly used libraries and the ones with the worst privacy behavior is a concerning privacy risk because the wide range of apps that use these libraries are already causing their users to reveal information that potentially enables persistent tracking.

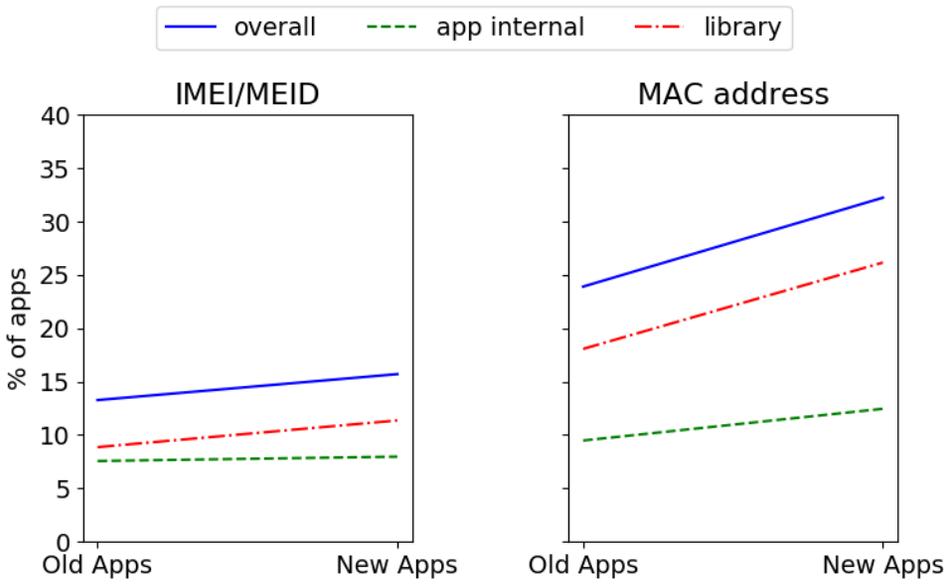

(a) Usage for specific hardware identifiers.

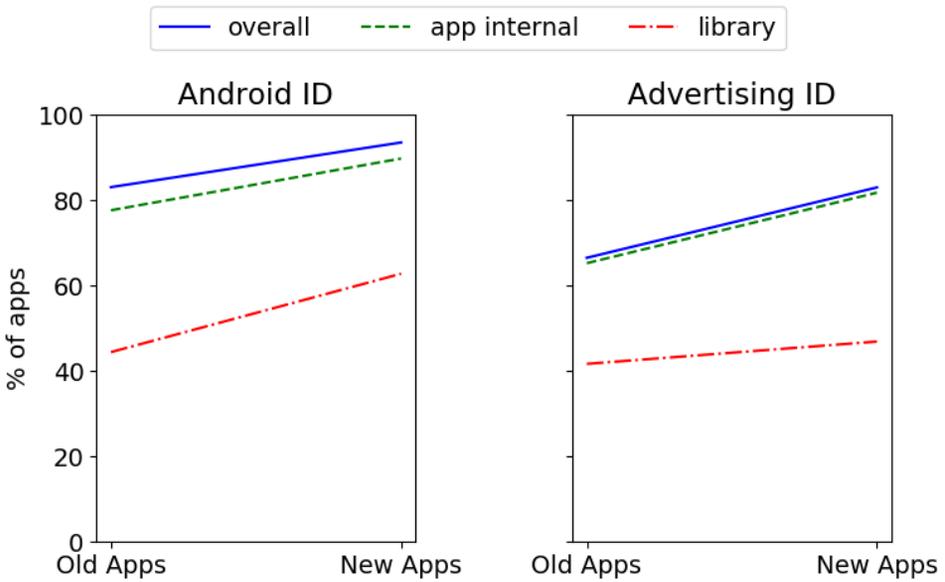

(b) Usage of identifiers that are not hardware-based

Fig. 17. When looking at the usage of some hardware-based identifiers, we observed that third-party libraries are more commonly the responsible party.

| Top 5 Most Commonly Used Libraries | | |
|---|---|---|
| **Library** | **# of ID Types** | **# of Hardware ID Types** |
| admob | 2 | 0 |
| GoogleAnalytics | 2 | 0 |
| firebase | 2 | 0 |
| facebook | 4 | 2 |
| squareup | 6 | 4 |
| **Top 5 Libraries with Most Privacy Issues** | | |
| **Library** | **# of ID Types** | **# of Hardware ID Types** |
| facebook | 4 | 2 |
| squareup | 6 | 4 |
| amazon | 8 | 6 |
| github | 8 | 6 |
| unity | 7 | 5 |

Table 7. This table shows how many types of identifiers for the top 5 most commonly used libraries and the 5 libraries with the most privacy issues, as well as how many hardware-based types. These common libraries use a lot of different types, with some using almost all of the types we analyzed for. This is concerning as it enables these libraries to track end users, especially with persistent hardware-based types.

## 9 DISCUSSION OF RESULTS

### 9.1 Changes Over Time

With regards to these specific aspects of privacy, there were somewhat mixed results about improvements over time. For example, the decrease in the number of apps that scanned for installed apps and in the number of unique identifiers used showed a definite improvement with time. The decrease in unique identifier usage was unexpected to us but can be attributed to developers becoming aware of better alternatives [21] and/or developing alternative business models that do not depend on ad tracking. On the other hand, there was no clear increase in HTTPS adoption, but there was no decrease either. We found this result particularly surprising given the launch of Let's Encrypt in 2016, which provides certificates for TLS encryption at no charge. However, this could be explained by third-party libraries adopting HTTPS at a slower rate, which we did observe as well.

While the increase in usage of sensitive permissions with increasing app maturity would seem to imply that privacy issues are getting worse, this heavily depends on the specific permissions being used and how necessary they are to the expressed functionality of the app. For example, the increase in usage of location data permissions could be due to app developers adding location-based functionality as a response to user comments or requests. This would corroborate previous work that surveyed developer commit history and found that, in most cases, permissions were added to an app to add functionality [36]. Studies determining if the use of a permission is justified have been conducted previously [5, 18, 25, 27, 38], and we leave the integration of these types of techniques with our data to future work.

These trends also lend supporting evidence to findings in past work that app developers tend to focus on implementing functionality and features first, putting privacy and security concerns as a lower priority. Loser et al. presented a case study of attempting to increase the motivation of mobile app developers to implement features required for security and privacy as this was commonly skipped over in agile development projects [23]. Other previous works conducted interviews and surveys to conclude that developers do indeed focus on privacy and security later on in the development of an app [7, 21]. Our work is the first to offer large-scale empirical data to support these previous qualitative findings.

When looking at the current state of privacy issues, the results are more negative. While HTTPS has been adopted for a majority of the flows in these apps, it is not as widespread as it should be. There is also quite a large percentage of apps in our data set that still scan for other installed apps on the device. However, these apps do not use a large number of sensitive permissions on average, and tend to use at most 1 persistent unique identifier. Furthermore, a very high percentage of them use advertising ID and/or Android ID, which are the better options when choosing a type of unique identifier to use.

Our conclusions here are limited by the fact that we were not able to analyze the APKs of every app on Google Play, but we were still able to gather enough supporting evidence from the ones that we did obtain.

## 9.2 Apps vs Libraries

When considering these specific aspects of privacy, third-party libraries are largely responsible for privacy issues exhibited by apps. These libraries have demonstrated that they are responsible for both a less widespread and a lower rate of adoption of HTTPS, and for driving growth in the number of apps that scan for other installed apps. Furthermore, third-party libraries used unique identifiers in a more egregious manner than the internal code of apps did, especially for types that are hardware-based and therefore more privacy-invasive.

Our investigation of the most commonly used libraries and the libraries that demonstrated the worst privacy behavior in these specific aspects also led to interesting observations. We found that HTTPS adoption was poor, especially among the most commonly used libraries. This further emphasizes the role libraries have played in slow HTTPS adoption; for example, AdMob was found in 92.0% of the apps we analyzed but only used HTTPS for 43.0% of links and 37.8% of domains. Almost all of these libraries made calls to at least one of the APIs for listing all the installed apps on a device. Furthermore, the libraries that demonstrated the worst overall privacy behavior used both sensitive permissions and unique identifiers egregiously, which is especially concerning given that some of these libraries were found in 35% to 42% of apps we analyzed.

Previous works have found that smartphone app developers are not always aware of the privacy risks associated with the use of certain given libraries, as well as how they used system APIs for data collection [21]. We believe that this can be greatly improved by enforcing third-party libraries have clearer and more publicly accessible privacy policies. In addition, more approachable and understandable guidelines around the usage of third-party libraries and how data can be collected via various APIs would be useful in helping improve the privacy issues associated with the usage of third-party libraries.

## 10 CONCLUSION

Through the framework we built for crawling and analyzing apps from Google Play, we have collected data on almost 1.9 million apps over the past 10 years. In addition, we have stored the APKs of 135,536 of those apps and used those APKs to conduct various static analyses to study general changes and trends in Android apps over this period of time. The trends we observed led us to a couple key findings that answered the questions of interest we initially set out to investigate. First, we found that privacy-related behavior improved as the amount of time an app had to undergo updates increased, corroborating the conclusions of previous work about privacy being a lower priority than functionality to developers. Second, we confirmed previous claims that third-party libraries are responsible for more issues with privacy than the internal code of apps, and offered new evidence to further support this.

We continue to use this framework to collect more and more data on Android apps as time goes on. In the future, we hope that this framework becomes a useful tool for other researchers or interested individuals to more easily conduct their own investigation on privacy within Android apps.

## A  FULL LIST OF SENSITIVE PERMISSIONS

| Permission name |
| --- |
| ACCESS_COARSE_LOCATION |
| ACCESS_FINE_LOCATION |
| ADD_VOICEMAIL |
| ANSWER_PHONE_CALLS |
| BODY_SENSORS |
| CALL_PHONE |
| CALL_PRIVILEGED |
| CAMERA |
| CAPTURE_AUDIO_HOTWORD |
| CAPTURE_AUDIO_OUTPUT |
| CAPTURE_SECURE_VIDEO_OUTPUT |
| CAPTURE_VIDEO_OUTPUT |
| DELETE_PACKAGES |
| GET_ACCOUNTS |
| LOCATION_HARDWARE |
| PROCESS_OUTGOING_CALLS |
| READ_CALENDAR |
| READ_CALL_LOG |
| READ_CONTACTS |
| READ_EXTERNAL_STORAGE |
| READ_LOGS |
| READ_PHONE_NUMBERS |
| READ_PHONE_STATE |
| READ_SMS |
| RECEIVE_MMS |
| RECEIVE_SMS |
| RECEIVE_WAP_PUSH |
| RECORD_AUDIO |
| SEND_SMS |
| USE_SIP |
| WRITE_CALENDAR |
| WRITE_CALL_LOG |
| WRITE_CONTACTS |
| WRITE_EXTERNAL_STORAGE |

Table 8.  This table lists all of the permissions we considered sensitive for the purposes of this paper. This list was informed by prior work done by the Pew Research Center [4] and by Google's published list of dangerous permissions [12].